\documentclass{appolb}
\usepackage{epsfig}
\begin{document}
\newcommand{\be}{\begin{equation}}
\newcommand{\ee}{\end{equation}}
\newcommand{\ba}{\begin{eqnarray}}
\newcommand{\tdm}[1]{\mbox{\boldmath $#1$}}
\def\qqd{(q\bar q)_{\rm dipole}}

\begin{titlepage}
\title{
Two-photon cross-sections from the saturation model
\footnote{Presented by L.~Motyka at the
Cracow Epiphany Conference on Quarks and Gluons in Extreme Conditions
                            3--6 January 2002, Cracow, Poland. }}

\author{J.\ Kwieci\'{n}ski$^{a}$,  L.\ Motyka$^{b,c}$ and N.\ T\^\i mneanu$^{b}$ \\[5mm]
{\small\it
$^{a}$ H.\ Niewodnicza\'{n}ski Institute of Nuclear Physics,
Krak\'{o}w, Poland \\ 
$^{b}$ High Energy Physics, Uppsala University, Uppsala, Sweden   \\
$^{c}$ Institute of Physics, Jagellonian University, Krak\'{o}w, Poland\\
}}

\maketitle

\begin{abstract}
A saturation model for the total $\gamma \gamma$
and $\gamma^{*} \gamma^{*}$ cross-sections and for the real photon structure
function $F_2^{\gamma}(x,Q^2)$ is described. The model is based on a QCD dipole 
picture of high energy scattering. The two-dipole cross-section is
assumed to satisfy the saturation property with the saturation radius taken
from the GBW analysis of the $\gamma^*p$ interaction at HERA. 
The model is combined with the QPM and non-pomeron reggeon contributions
an it gives a very good description of the data on
the $\gamma \gamma$ total cross-section, on the photon structure function
$F_2^{\gamma}(x,Q^2)$ at low $x$ and on the $\gamma^* \gamma^*$
cross-section. Production of heavy quarks in $\gamma \gamma$ collisions is 
also studied. 
\end{abstract}
\end{titlepage}

\section{Introduction}

The saturation model \cite{GBW} was proven to provide a very
efficient framework to describe variety of experimental results
on high energy scattering. With a very small number of 
free parameters, Golec-Biernat and W\"{u}sthoff (GBW) fitted 
low~$x$ data from HERA for both inclusive
and diffractive scattering \cite{GBW}. 
Some promising results were also obtained for elastic
vector meson photo- and electroproduction \cite{VMSAT}. 

The central concept behind the saturation model is 
an $x$~dependent saturation scale $Q_s(x)$ at which
unitarity corrections to the linear parton evolution
in the proton become significant. 
In other words, $Q_s(x)$ is a typical scale of a hard probe 
at which a transition from a single scattering to a multiple 
scattering regime occurs.

The model is well grounded in perturbative QCD.
The existence of such a scale in the
saturation domain was suggested already 
in \cite{BL} as a consequence of the GLR
equation \cite{GLR} obtained in the double
logarithmic approximation. 
A parton evolution equation involving unitarity
corrections at LL-$1/x$ approximation and the large-$N_c$ limit
was derived by Balitsky and Kovchegov (BK) \cite{BK}.
Numerous studies \cite{BKsol}
showed that the solutions to the BK equation are, 
with a good approximation, consistent with the    
presence of the saturation scale.

Our idea was to extend the saturation model constructed
for $\gamma^*p$ scattering to describe also 
$\gamma^* \gamma^*$ cross sections.  
The successful extension, performed in \cite{TKM},
provided a test of the saturation model in a new
environment and confirmed the universality of the model.  
Results obtained in \cite{TKM} are also of some importance 
for two-photon physics, since the model is capable of 
describing a broad set of observables in wide kinematical
range in a simple, unified framework. 
In this presentation the most important results of \cite{TKM}
will be summarized.


\section{The model}

The saturation model for two-photon interactions
is constructed in analogy to the GBW model \cite{GBW}. 
In terms of the virtual photon four-momenta $q_1$ and $q_2$ 
we have $Q^2 _{1,2}=-q_{1,2}^2$ and $W^2 = (q_1+q_2)^2$, 
see Fig.~\ref{diagram}.
Each of the virtual photons is decomposed into colour dipoles
$\qqd$ representing virtual components of the photon in the
transverse plane and their distribution in the photon is assumed to 
follow from the perturbative formalism.

\begin{figure}[t]
\begin{center}
\epsfig{width= 0.5\columnwidth,file=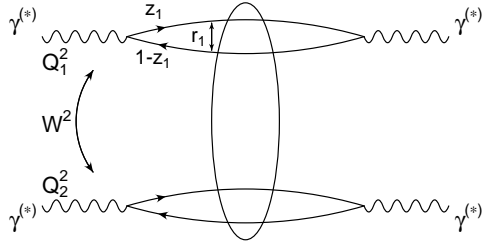}
\caption{\small\it The diagram illustrating the $\gamma^* \gamma^*$ interaction in the
dipole representation}
\label{diagram}
\end{center}
\end{figure}

A formula for the two-photon cross-section part coming from the exchange of
{\em gluonic} degrees of freedom reads \cite{DDR}
\[
\sigma^G_{ij}(W^2,Q_1^2,Q_2^2)\; = \;
\]
\be
\sum_{a,b=1}^{N_f} \int_0^1dz_1\int d^2 {\tdm r_1}|\Psi_i^a(z_1,{\tdm r_1})|^2
\int_0^1 dz_2\int d^2 {\tdm r_2}|\Psi_j^b(z_2,{\tdm r_2})|^2
\; \sigma^{dd}_{a,b}(\bar x_{ab},r_1,r_2) ,
\label{master}
\end{equation}
where
the indices $i,j$ label the polarisation states of the virtual photons,
i.e.\ $T$ or $L$ and $\sigma^{dd}_{a,b}(\bar x_{ab},r_1,r_2)$ are the
dipole-dipole total cross-sections corresponding to their
different flavour content specified by $a$ and $b$.
The transverse vectors ${\tdm r}_k$ denote the separation between 
$q$ and $\bar q$ in the colour dipoles and $z_k$ are the longitudinal 
momentum fractions of the quark in the photon~$k$ ($k=1,2$).
The photon wave functions are given by 
\[
|\Psi_{T}^a (z,{\tdm r})|^2 = {6\alpha_{em}\over 4 \pi^2} e_a^2
\{ [z^2+(1-z)^2]\;\epsilon_a ^2 K_1^2 (\epsilon_{a}r)
+m_f^2\,K_0^2(\epsilon_{a}r) \}
\]
\be
|\Psi_{L}^a (z,{\tdm r})|^2 \; = \;
{6\alpha_{em}\over\pi^2} e_a^2
 Q^2 z^2 (1-z)^2 \; K_0^2 (\epsilon_{a}r), 
\label{psit}
\end{equation}
with
\begin{equation}
{(\epsilon_a ^k)}^2 \; = \; z_k(1-z_k)Q^2+m_a^2,\qquad k=1,2,
\label{epsilon}
\end{equation}
where $e_a$ and $m_a$ denote the charge and mass of the quark of flavour $a$.
The functions $K_0$ and $K_1$ are the McDonald--Bessel functions.

Inspired by the GBW simple choice for the dipole-proton cross-section,
we use the following parametrisation of the dipole-dipole cross-section
$\sigma_{a,b}$
\begin{equation}
\sigma^{dd}_{a,b}(\bar x_{ab},r_1,r_2)\; = \;\sigma_0^{a,b}\left[
1- \exp\left(-{r_{\rm eff}^2\over  4R_0^2(\bar x_{ab})}\right)
\right],
\label{sigmadd}
\end{equation}
where for $\bar x_{ab}$ we take the following expression symmetric in~$(1,2)$
\begin{equation}
\bar x_{ab} \; = \;{Q_1^2 + Q_2^2 +4m_a^2+4m_b^2\over W^2+Q_1^2+Q_2^2},
\label{barx}
\end{equation}
which allows an extension of the model down to the limit $Q_{1,2}^2=0$.
Note, that $\bar x_{ab}$  depends on the flavour of scattering quarks.
We use the same parametrisation of the saturation radius $R_0(\bar x)$
as that in equation (7) in \cite{GBW}, i.e.\
\begin{equation}
R_0(\bar x)\; = \;{1\over Q_0} \left({\bar x\over x_0}\right)^{\lambda/2},
\label{r0}
\end{equation}
and adopt the same set of parameters defining this quantity as those in
\cite{GBW}. For the saturation value $\sigma_0^{a,b}$
of the dipole-dipole cross-section (cf.\ equation (\ref{sigmadd})) we set
\begin{equation}
\sigma_0^{a,b}\; = \;{2\over 3}\sigma_0,
\label{sigma0}
\end{equation}
where $\sigma_0$ is the same as that in \cite{GBW}.
For light flavours, equation (\ref{sigma0}) can be justified by 
the quark counting rule, as the ratio between  the number of 
constituent quarks in a photon  and the
corresponding number of constituent quarks in the proton.
We also use the same value of 
$\sigma_0^{a,b}$ for all flavours.

Three scenarios for $r_{\rm eff}(r_1,r_2)$ are considered:
\begin{enumerate}
\item $\displaystyle r^2_{\rm  eff}\; = \;{r_1^2r_2^2\over r_1^2+r_2^2}$,
\item $r^2_{\rm eff}\; = \;\min(r_1^2,r_2^2)$,
\item $r^2_{\rm  eff}\; = \;\min(r_1^2,r_2^2)[1+\ln(\max(r_1,r_2)/\min(r_1,r_2))]$.
\end{enumerate}
All three parametrisations exhibit colour 
transparency, i.e.\ \\
$\sigma^{dd}_{a,b}(\bar x,r_1,r_2) 
\rightarrow~0$ for  $r_{1} \rightarrow 0$ or $r_{2} \rightarrow 0$.
Cases (1) and (2) reduce to the original GBW model
when one of the dipoles is much larger than the other
and  option (3), being significantly different from
(1) and (2), is a controll case.


The saturation model accounts for an exchange of {\em gluonic} degrees of
freedom, the QCD pomeron fan diagrams. Such exchanges dominate
at very high energies (low~$x$) but at lower energies the processes
involving quark exchange have to be considered as well. 
Thus, in order to get a complete description of $\gamma^* \gamma^*$ interactions
 we should add to the `pomeron' contribution defined by equation (\ref{master}) 
the non-pomeron reggeon and QPM terms \cite{JKLM}. 
The additional contributions are characterised by a decreasing
energy dependence, i.e.\ $\sim 1/W^{2\eta}$ for the reggeon and 
$\sim 1/W^2$ (with $\ln W$ corrections) for QPM.  
The QPM contribution, represented by the quark box
diagrams, is well known and the cross-sections are given, for instance,
in \cite{BUD}.
The reggeon contribution represents a non-perturbative phenomenon
related to Regge trajectories of light mesons.
It is known mainly from fits to total hadronic cross-sections and to the proton
structure function $F_2$. We used the following  parametrisation of the reggeon 
exchange cross-section in two-photon interactions \cite{DDR}
\be
\sigma^R(W^2,Q_1^2,Q_2^2)\; = \; 4\pi^2\alpha_{em}^2\frac{A_2}{a_2}\left[{a_2^2\over
(a_2+Q_1^2)(a_2+Q_2^2)}\right]^{1-\eta}\left({W^2\over a_2}\right)^{-\eta}.
\label{reggeon}
\ee
We have chosen $\eta=0.3$ in accordance with the 
value of the Regge intercept of the $f_2$ meson trajectory
$1-\eta=0.7$ \cite{PVLF0}. Parameters $A_2$ and $a_2$ 
were fitted to the data on two-photon collisions. 

Formulae (\ref{master}) and (\ref{reggeon}) 
describing the gluonic and reggeon components are  
valid at asymptotically high energies, where
the impact of kinametical thresholds is small.
The threshold effects are approximately
accounted for by introducing a multiplicative 
correction factors, whose form is deduced
form spectator counting rules (see \cite{TKM}).

Thus, the total $\gamma^*(Q_1^2)\gamma^*(Q_2^2)$ cross-section 
reads
\be
\sigma_{ij} ^{\rm tot} \;=\;
\tilde\sigma_{ij} ^G  + 
\tilde\sigma ^R \delta_{iT} \delta_{jT}+
\sigma_{ij} ^{\rm QPM},
\label{total}
\ee
where
$
\tilde\sigma_{ij} ^G (W^2,Q_1^2,Q_2^2) 
$
is the gluonic component, corresponding to dipole-dipole scattering,
as in eq.\ (\ref{master}), but with the dipole-dipole cross-section
including the threshold correction factor 
\be
\tilde\sigma^{dd}_{a,b}(\bar x_{ab},r_1,r_2) \; = \;
     (1-\bar x_{ab})^5\, \sigma^{dd}_{a,b}(\bar x_{ab},r_1,r_2), 
\label{sigmaddth}
\ee
c.f.\ eq.\ (\ref{sigmadd}), and $\bar x_{ab}$ is given
by eq.\ (\ref{barx}).
The sub-leading reggeon contributes only to scattering of two 
transversely polarised photons and also contains a threshold
correction
\be
\tilde\sigma ^R (W^2,Q_1^2,Q_2^2) = (1-\bar{x})\, \sigma ^R (W^2,Q_1^2,Q_2^2),
\ee  
with 
\be
\bar x = {Q_1^2 + Q_2 ^2 + 8 m_q ^2 \over W^2 + Q_1^2 + Q_2 ^2}.
\ee
The third term $\sigma_{i,j} ^{\rm QPM} (W^2,Q_1^2,Q_2^2)$
is the standard QPM contribution.

\section{Comparison to experimental data}

\subsection{Parameters of models}

In the comparison to the data we study three models, based on all cases
for the effective radius, as described in Section~2.2. We will refer to 
these models as Model 1, 2 and 3, corresponding to the choice of the
dipole-dipole cross-section.
Let us recall that we take without any modification the parameters
of the GBW model: $\sigma_0 = 29.13 $~mb, $x_0= 0.41 \cdot 10^{-4}$ and  $\lambda= 0.277$.
However, we fit the light quark mass to the two-photon data, since it is not
very well constrained by the GBW fit, as we explicitly verified.
On the other hand, the sensitivity of the choice
of the mass appears to be large for the two-photon total cross-section.
We find that the optimal values of the light quark ($u$, $d$ and $s$) masses
$m_q$ are 0.21, 0.23 and 0.30~GeV in Model 1,~2 and~3 correspondingly.
Also, the masses of the charm and bottom quark are tuned within the range
allowed by current measurements, to get the  optimal global description
in Model~1, $\displaystyle r^2_{\rm  eff}\; = {r_1^2r_2^2 / (r_1^2+r_2^2)}$,
which agrees best with data. For the charm quark we use $m_c = 1.3$~GeV
and for bottom $m_b = 4.5$~GeV.
The values of parameters in the reggeon term (\ref{reggeon}): $\eta = 0.3$, 
$A_2 = 0.26$ and $a_2 = 0.2\;{\rm GeV}^2$ are found to give the best description 
of data, when combined with the saturation model.
The values of masses listed above are consistently used also in the quark
box contribution (QPM). The Models, which we shall mention from now on, 
contain the saturation models described in Section 2, combined with
the reggeon and QPM contribution.

The references to the relevant  experimental papers may be found
in \cite{TKM}.

\subsection{The test case: the $\gamma p$ total cross-section}

\begin{figure}[t]
\begin{center}
\epsfig{width= 0.6\columnwidth,file=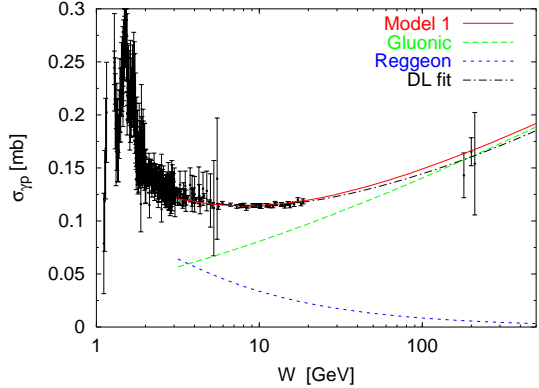}
\caption{\small\it The total $\gamma p$ cross-section -- predictions from the GBW model with
the light quark mass $m_q$ set to $0.21$~GeV and the charmed quark mass $m_c =1.3$~GeV,
supplemented by the reggeon term (\ref{regggp}), compared to data and
to the Donnachie-Landshoff fit.
}

\label{realgp}
\end{center}
\end{figure}

In order to describe two photon data, we altered the
original light quark mass of the GBW model.
Besides that, we included the reggeon term and 
the threshold correction factors in
the analysis. Thus, it is worthwile to compare
the results from the modified model with 
the data on the $\gamma p$ total cross-section.
Thus we calculated the dipole-proton scattering contribution
using the original GBW approach, with the light quark mass, $m_q$, set
to 0.21~GeV, as in Model 1, and added the reggeon term
\be
\sigma_{\gamma p} ^R (W^2) \; = \;
A_{\gamma p}\; \left( {W^2 \over 1\,{\rm GeV}^2} \right) ^{-\eta} ,
\label{regggp}
\ee
where $A_{\gamma p}$ was fitted to data and the best value reads 
$A_{\gamma p}=0.135~{\rm mb}$.
The result is given in Fig.~\ref{realgp}, where the cross-section
from Model~1 is compared to the experimental data and to the classical
Donnachie-Landshoff fit \cite{DLgp}.
The fitted curve, with only one free parameter  $A_{\gamma p}$ follows the
data accurately, suggesting that the model has certain universal properties.

\subsection{Total $\gamma\gamma$ cross-section}

\begin{figure}[t]
\begin{center}
\epsfig{width= 0.6\columnwidth,file=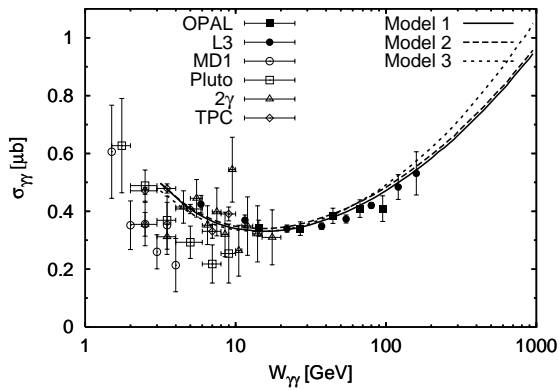} \hspace{4em} \\
\caption{\small\it The total $\gamma\gamma$ cross-section: 
data compared with predictions from all three Models.}
\label{real-all}
\end{center}
\end{figure}

The available data for the $\gamma\gamma$ total cross-section range
from the $\gamma\gamma$  energy $W$ equal to about 1~GeV up to
about 160~GeV, 
see Fig.\ \ref{real-all}.
The experimental errors of the data are, unfortunately, rather large.
One of the reasons is that those data were taken for virtual photons
coming from electron beams and then the results were extrapolated to zero 
virtualities. Some uncertainty is caused by the reconstruction
of actual $\gamma\gamma$ collision energy from the visible
hadronic energy. In such a reconstruction one relies on an
unfolding procedure, based on a Monte Carlo program. 
In Fig.~\ref{real-all} we show the total $\gamma\gamma$ cross-section 
from the Models, obtained using eq.\ (\ref{total}) with $i=j=T$.
The data from LEP were unfolded with \textsc{Phojet}.
The agreement with data is very good down to $W \simeq 3$~GeV
for all the Models.

\begin{figure}[t]
\begin{center}
\begin{tabular}{cc}
\epsfig{width= 0.45\columnwidth,file=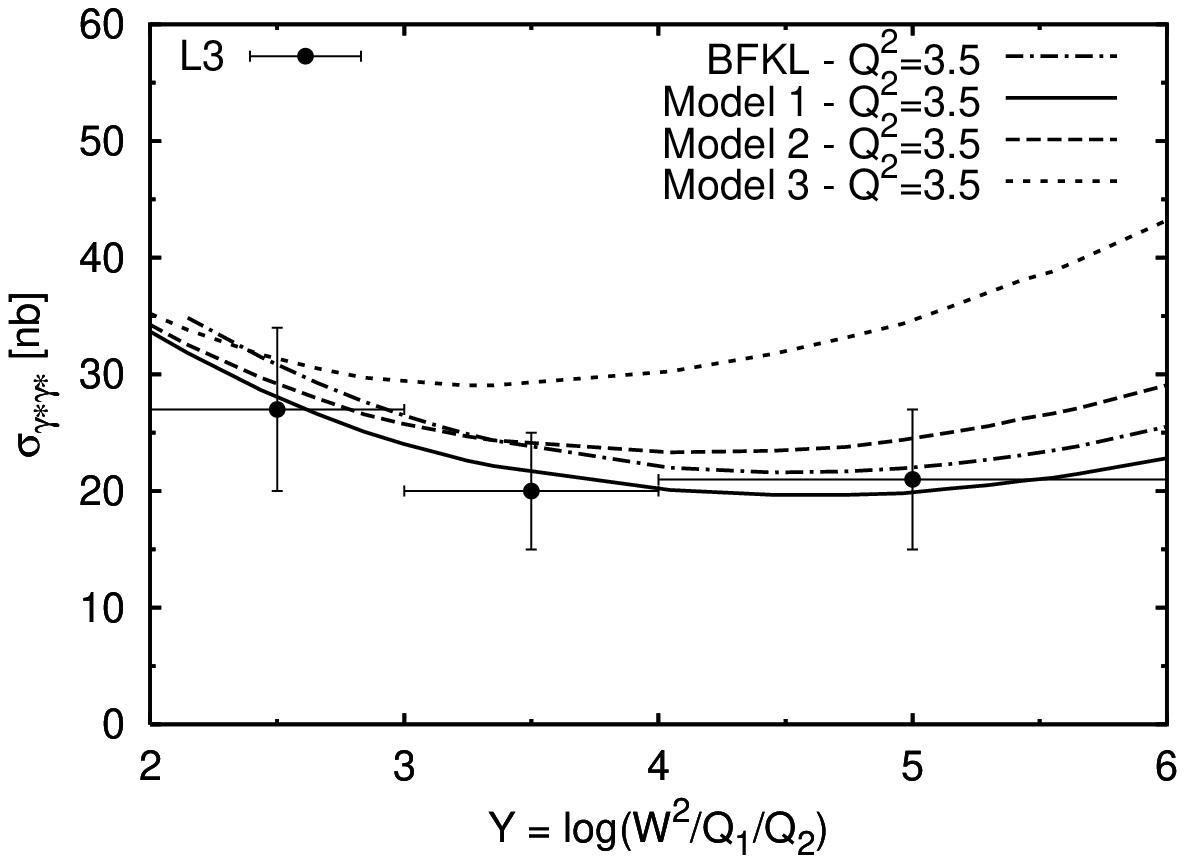} &
\epsfig{width= 0.45\columnwidth,file=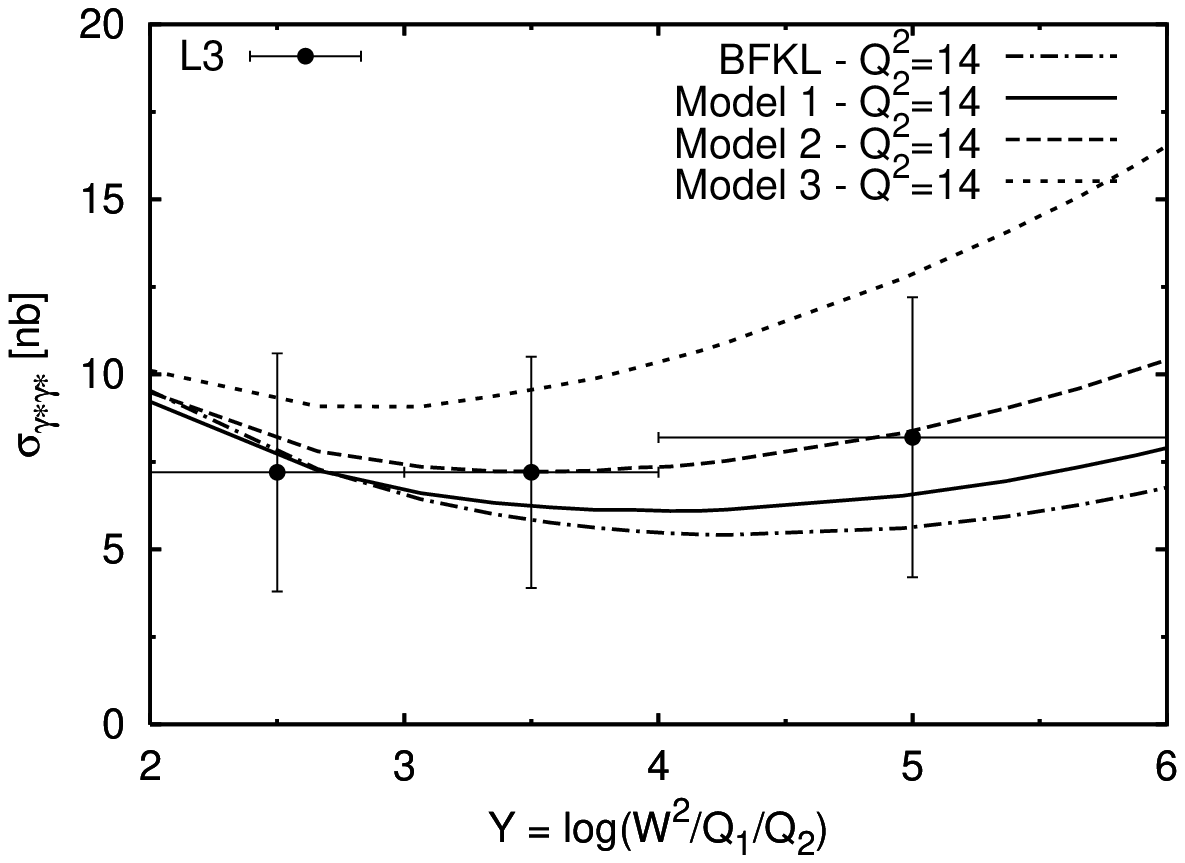}\\
a) & b) \\
\end{tabular}
\epsfig{width= 0.45\columnwidth,file=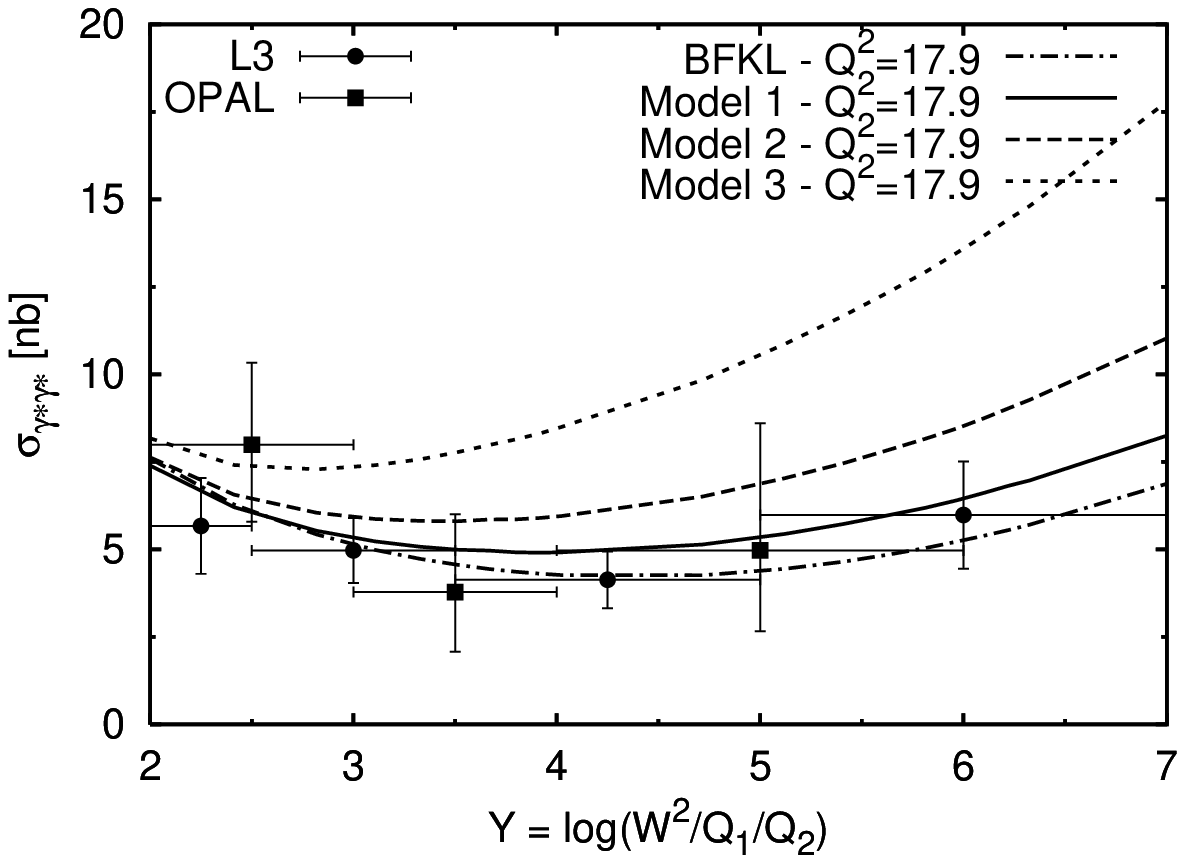} \\
c) \\
\caption{\small\it
Total $\gamma^*\gamma^*$ cross-section for
            (a) $Q ^2= 3.5~{\rm GeV}^2$,
            (b) $Q^2 = 14~{\rm GeV}^2$ and
            (c) $Q ^2=17.9~{\rm GeV}^2$ -- comparison between LEP data 
            and the Models plotted as a function of $Y= \ln (W^2/Q^2)$.
            Also shown is the result of Ref.~\cite{JKLM} based on the
            BFKL formalism with subleading corrections, supplemented by
            the QPM term, the soft pomeron and the subleading reggeon
            contributions.}
\label{virt-all}
\end{center}
\end{figure}

\subsection{Total $\gamma^*\gamma^*$ cross-section}

The data for the total $\gamma^*\gamma^*$ cross-section are extracted
from so-called double-tagged events, that is from $e^+e^-$ events
in which both the scattered electrons are measured and hadrons are produced.
In such events  measurement of the kinematical variables of the leptons
determines both the virtualities $Q_1^2$ and $Q_2^2$ of the colliding photons and
the collision energy $W$. The tagging angles in LEP experiments restrict the
virtualities to be similar, i.e $Q_1^2 \sim Q_2^2 = Q^2$.
The data are available from LEP for average values
$Q ^2=3.5~{\rm GeV}^2$, $14~{\rm GeV}^2$ and $Q ^2=17.9~{\rm GeV}^2$
in a wide range of $W$.

In Figs.~\ref{virt-all}a,b,c those data are compared with the curves from 
the Models. As an estimate of the total $\gamma^*\gamma^*$ cross-section 
we use a simple sum of the cross-sections $\sigma_{ij} ^{\rm tot}$ 
(eq.\ (\ref{total})) over transverse and longitudinal polarisations $i$ and~$j$
of both photons. 
In addition we plot also the prediction obtained in Ref.~\cite{JKLM} by solving
the BFKL equation with non-leading effects, and added phenomenological
soft pomeron and reggeon contributions and the QPM term.
Models 1 and~2 fit the data well whereas Model~3 does not.

The virtuality of both photons are large, so the unitarity corrections, 
the light quark mass effects and the reggeon contribution are not 
important here. Moreover, the perturbative approximation for the photon 
wave function is fully justified in this case. 
Thus, in this measurement the form of the dipole-dipole
cross-section is directly probed.

\begin{figure}[t]
\begin{center}
\begin{tabular}{cc}
\epsfig{width= 0.45\columnwidth,file=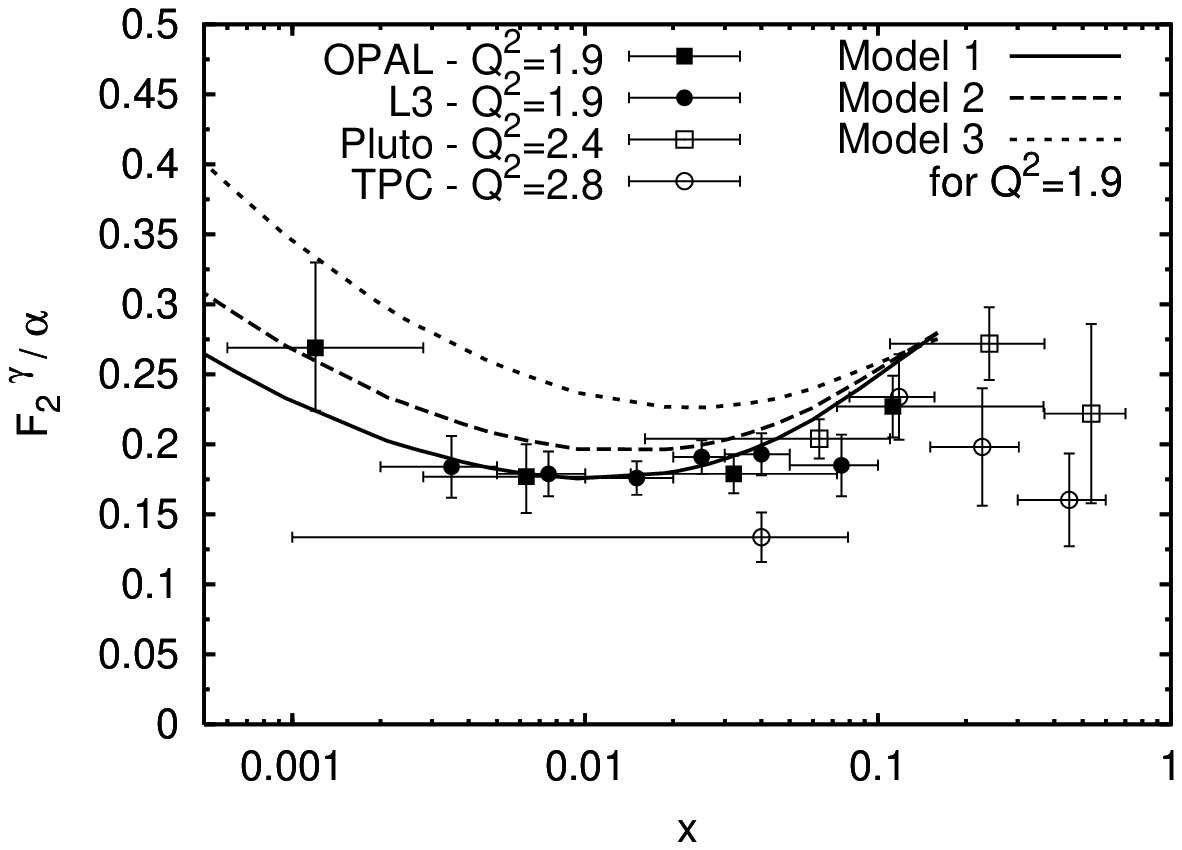} &
\epsfig{width= 0.45\columnwidth,file=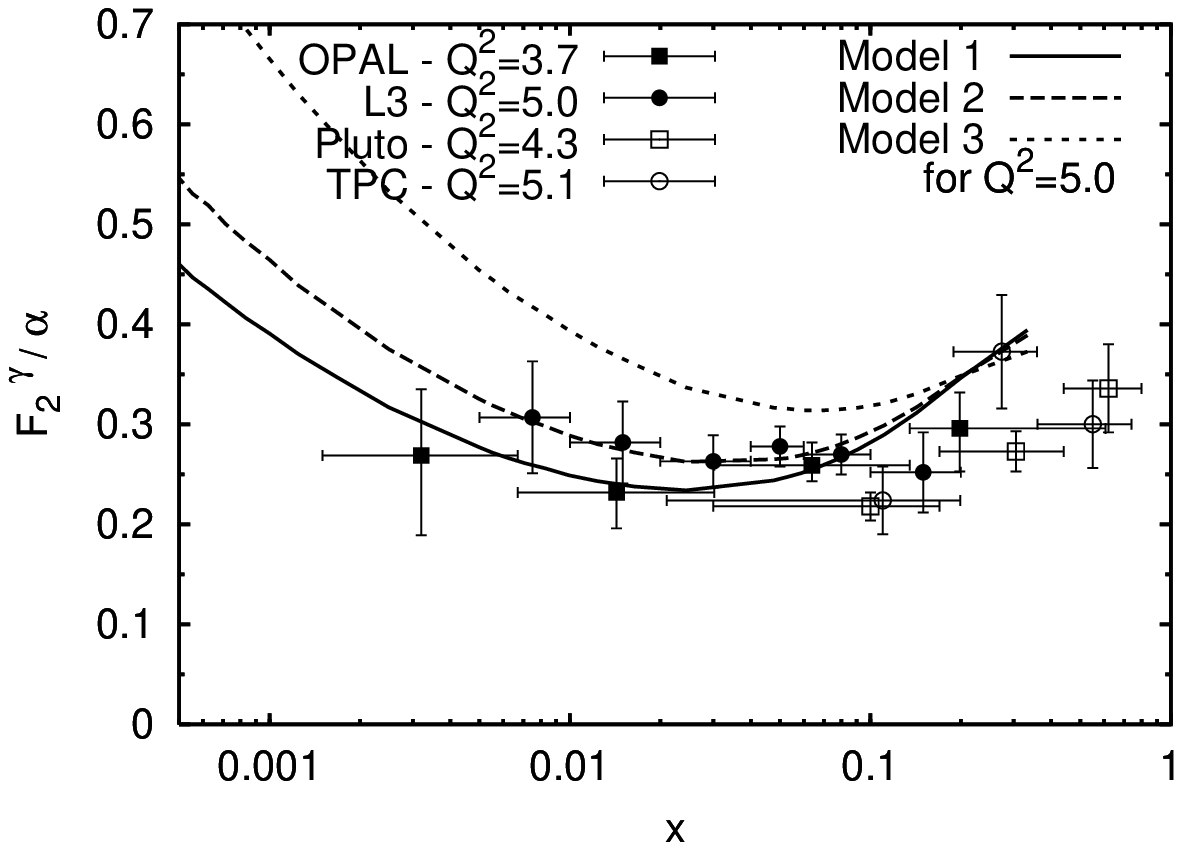}\\
a) & b) \\
\epsfig{width= 0.45\columnwidth,file=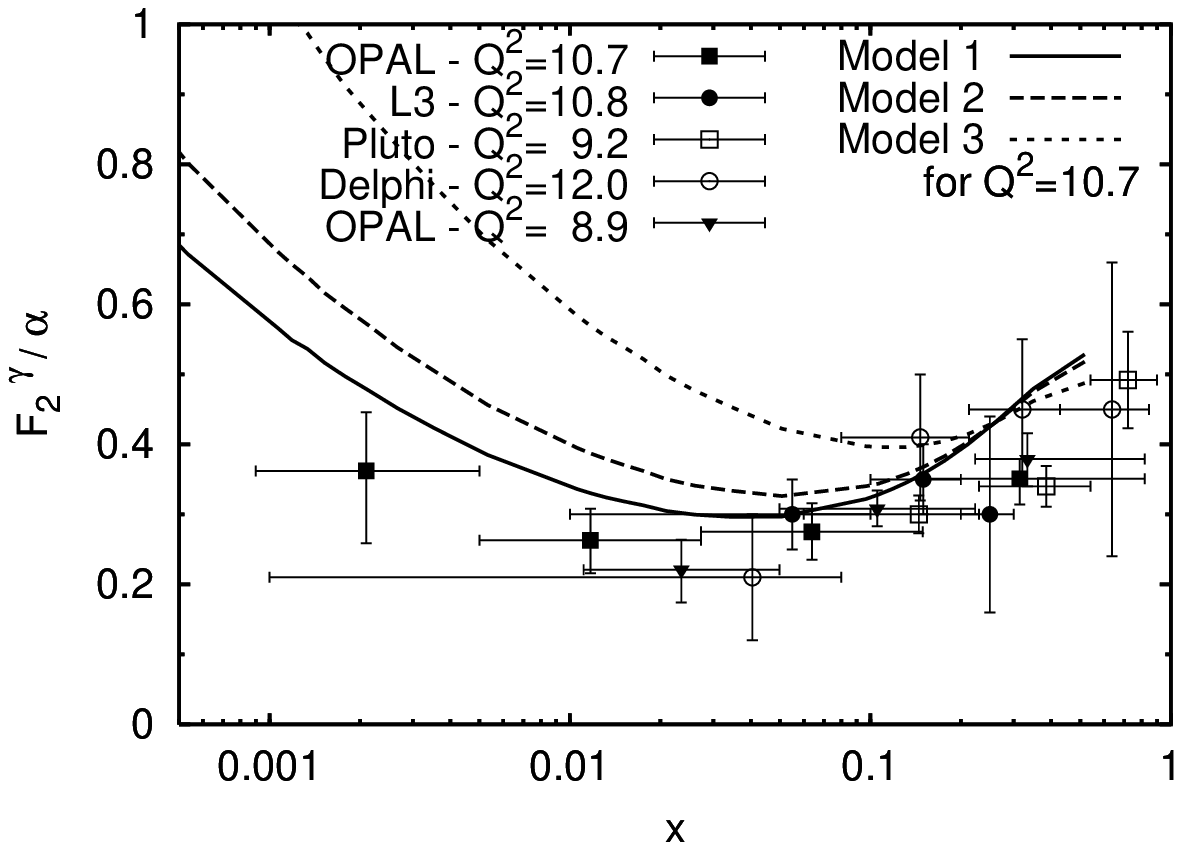} &
\epsfig{width= 0.45\columnwidth,file=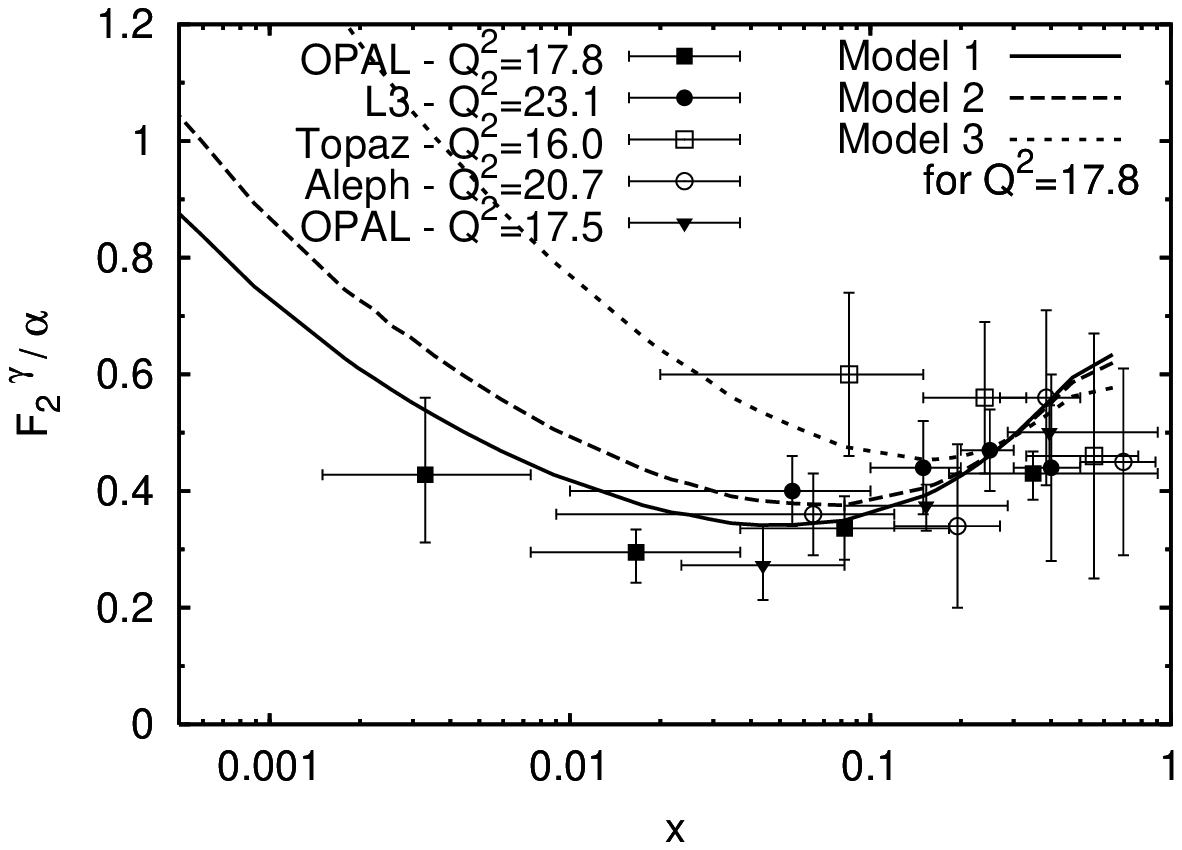} \\
c) & d) \\
\end{tabular}
\caption{\small\it The photon structure function $F_2 ^{\gamma}(x,Q^2)$:
the experimental data compared to predictions following from the
Models for various $Q^2$: (a) from 1.9 to 2.8~GeV$^2$, (b) from 3.7 to 5.1~GeV$^2$,
(c) from 8.9 to 12.0~GeV$^2$ and (d) from 16.0 to 23.1~GeV$^2$.}
\label{f2-all}
\end{center}
\end{figure}

\subsection{Photon structure}

The data on quasi-real photon structure are obtained 
mostly in single tagged $e^+e^-$ events, in which
a two-photon collision occurs. One of the photons has 
a large virtuality and probes the other, almost real photon.
In Fig.~\ref{f2-all}  we show  the comparison of our predictions with the 
experimental data 
for the  virtuality $Q^2$ in the range from
(a)~1.9 to~2.8~GeV$^2$,
(b)~3.7 to~5.1~GeV$^2$,
(c)~8.9 to~12.0~GeV$^2$
and finally (d)~from 16.0 to~23.1~GeV$^2$.
Note, that in each plot the data for various virtualities are
combined. In each plot the value of virtuality $Q^2$ adopted to obtain
the theoretical curve is indicated and was selected to match the average 
value $Q^2$ of the data-set containing the best data at low $x$. 
Model 1, favoured by the $\gamma^*\gamma^*$ data provides the best description
of $F_2 ^{\gamma}$ as well. 

\subsection{Heavy flavour production}

Another interesting process which we have studied in the dipole model
is the production of heavy flavours (charm and bottom) in $\gamma\gamma$
collisions. Heavy quarks can be produced by three mechanisms:
a direct production, a direct photoproduction off a resolved photon 
and a process with two resolved photons. The last mechanism
is not accounted for in our approach.  

The reggeon exchange is a non-perturbative phenomenon and should not
contribute to heavy flavour production, so it is assumed to vanish here.
In Fig.~\ref{charm} we plot the predictions from all three Models compared 
with L3 data on charm production. The best model, Model~1, is slightly 
below the data. The shape of the cross-section  is well reproduced.

\begin{figure}[t]
\begin{center}
\begin{tabular}{cc}
\epsfig{width= 0.45\columnwidth,file=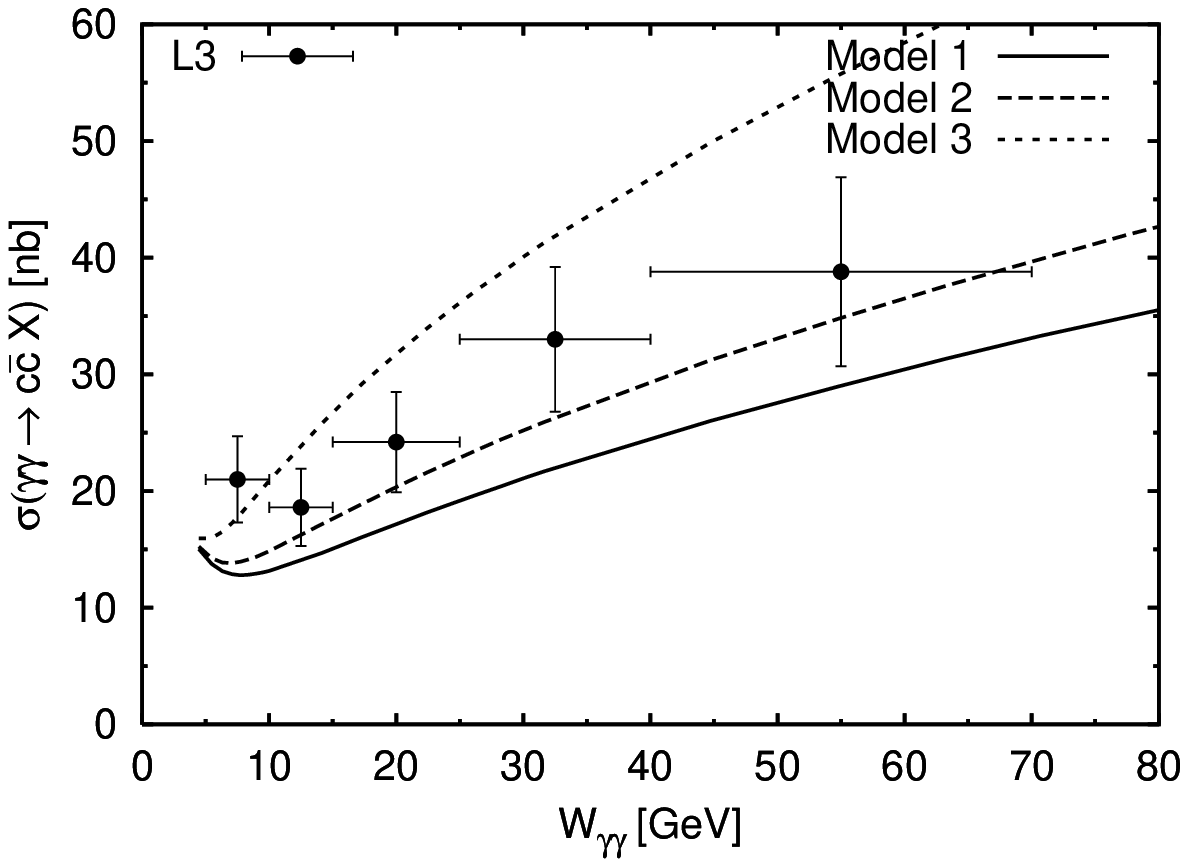} &
\epsfig{width= 0.45\columnwidth,file=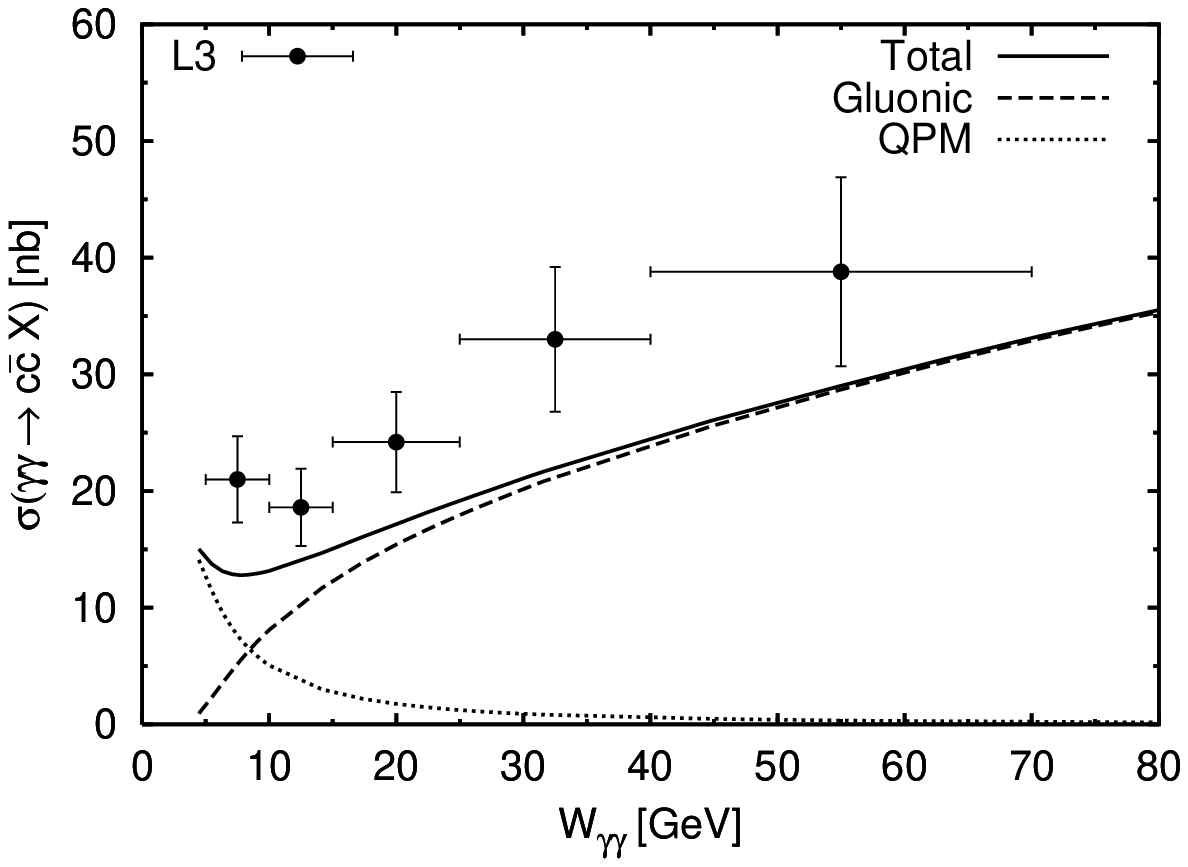} \\
a) & b) \\
\end{tabular}
\caption{\small\it The cross section for the inclusive charm production in $\gamma\gamma$
collisions: (a) results for all three Models and (b) the decomposition of
the result from Model~1 on the QPM and gluonic component.}
\label{charm}
\end{center}
\end{figure}

Production of bottom quarks in two almost real photon collisions was
investigated experimentally by the L3 
and the OPAL 
collaborations. There, the measured process was
$e^+e^- \to e^+e^- b\bar{b} X$, with anti-tagged electrons
at $e^+e^-$ invariant collision energies $\sqrt{s}_{ee}$ between 189~GeV and 202~GeV.
The total cross-section for this reaction was found to be
$13.1 \pm 2.0 \, {\rm (stat)} \pm 2.4 \,{\rm (syst)}$~pb (L3) and
$14.2 \pm 2.5 \, {\rm (stat)} \pm 5 \,{\rm (syst)}$~pb (OPAL)
whereas the theoretical estimate from Model~1  for $\sqrt{s}_{ee} = 200$~GeV
gives about 5.5~pb with less than 10\% uncertainty related to the choice of $b$-quark mass.
This is significantly below the experimental data but above
the expectations of $3 \pm 1$~pb based on standard QCD calculations 
with the use of the resolved photon approximation.

In conclusion, the saturation model underestimates the cross-section
for production of heavy quarks and the discrepancy increases with
increasing quark mass, or perhaps, decreasing electric charge.

\section{Conclusions}

In this contribution an extension of the saturation approach 
to two photon physics has been presented.
This extension required an explicit model for the scattering of
two colour dipoles. We considered three models of this cross-section,
all of them exhibiting the essential feature of colour transparency
for small dipoles, and the saturation property for large ones.
We kept the GBW form of the unitarising function and the original
parameters, except for changing the values of quark masses, which 
was necessary to describe the data on the  total two real photon 
cross-section. 
In order to obtain a more complete description applicable at lower energies
the saturation model has been combined with other, well known contributions  
related to the quark box diagram and non-pomeron reggeon exchange. 

Our theoretical results were compared with the data for
different two-photon processes at high rapidity values: the total
$\gamma\gamma$ cross-section, the total $\gamma^*\gamma^*$ cross-section
for similar virtualities of the photons, the real photon structure
function $F_2 ^{\gamma}$ and heavy flavour production. Free parameters
were fitted to the data.  With the best model a reasonable global 
description of the available two-photon data was obtained, 
except for the $b$-quark production.
Thus, the saturation model was found to provide a simple and efficient
framework to calculate observables in two-photon processes.

\section*{Acknowledgments}
We gratefully acknowledge the support from the Swedish Natural
Science ResearchCouncil and  by the Polish Committee for Scientific 
Research (KBN) grants no.\ 2P03B~05119 and 5P03B~14420.

\end{document}